\documentclass{PoS}

\pdfoutput=1

\usepackage{amsmath}
\usepackage{amstext}
\usepackage{amsfonts}
\usepackage{amssymb}

\title{\vspace*{-2cm}{\large \normalfont \hfill DESY 10-190, JLAB-THY-10-1287, SFB/CPP-10-101}\\\vspace*{1cm}
Nucleon matrix elements with $N_f=2+1+1$ maximally twisted fermions\vspace*{0.3cm}
\begin{center}
 \includegraphics[scale=0.2]{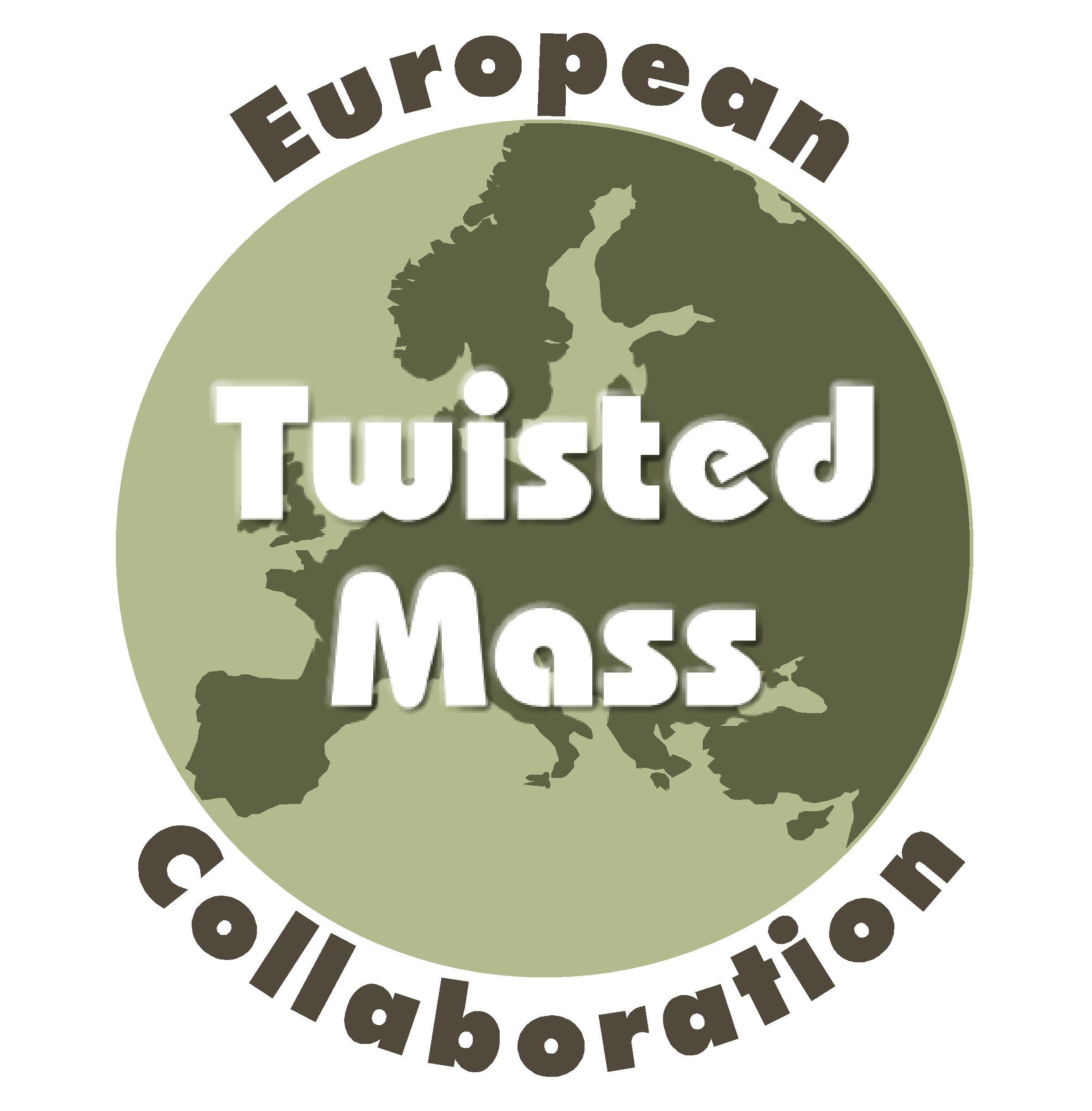}
\end{center}
}

\ShortTitle{Nucleon matrix elements with maximally twisted fermions}

\author{
\speaker{Simon Dinter}$\,^a$,
Constantia Alexandrou$\,^{b,c}$,
Martha Constantinou$\,^b$,
Vincent Drach$\,^a$,
Karl Jansen$\,^a$ and
Dru Renner$\,^a$\thanks{Current address: Jefferson Lab.}\vspace{0mm}\\
\llap{$^a$}{NIC, DESY Zeuthen, Platanenallee 6, D-15738 Zeuthen, Germany\\}
\llap{$^b$}{Department of Physics, University of Cyprus, P.O. Box 20537, 1678 Nicosia, Cyprus\\}
\llap{$^c$}{Computation-based Science and Technology Research Center,
The Cyprus Institute, 15 Kypranoros Str., 1645 Nicosia, Cyprus\\}

E-mail: \email{Simon.Dinter@desy.de}
}

\abstract{We present the first lattice calculation of nucleon matrix elements
using four dynamical flavors.  We use the $N_f=2+1+1$ maximally twisted 
mass formulation. The renormalization 
is performed non-perturbatively in the RI$^\prime$-MOM scheme and results
are given for the vector and axial vector operators with up to
one-derivative.  Our calculation of the average momentum of the unpolarized
non-singlet parton distribution is presented and compared to our previous results
obtained from the $N_f=2$  case.}

\FullConference{The XXVIII International Symposium on Lattice Field Theory\\
                June 14-19,2010\\
                Villasimius, Sardinia Italy}

\begin{document}
\section{Introduction}
The European Twisted Mass Collaboration (ETMC) is 
generating gauge configurations with four dynamical flavors:\ two degenerate light quarks
and a pair of strange and charm quarks with their masses approximately fixed to their physical values
($N_f=2+1+1$). Several volumes and
 lattice spacings smaller than 0.1~fm are being produced~\cite{Baron:2010bv,Baron:2009zq}. We present here first results using the $N_f=2+1+1$ formulation
 to calculate
observables probing nucleon structure.
In particular we show results for
the non-singlet moment of the nucleon's unpolarized parton distribution
$\left\langle x \right\rangle_{u-d}$.
This is the lowest non-trivial moment
of the quark momentum distribution. 
There are no disconnected contributions, which are computationally very
demanding, and hence $\left\langle x \right\rangle_{u-d}$ 
can be computed relatively easily with lattice QCD.
Furthermore, $\left\langle x \right\rangle_{u-d}$ is also 
known accurately from global analyses of experimental measurements,
thus it provides a good benchmark for lattice computations of nucleon structure.
The successful determination of $\langle x\rangle_{u-d}$ will provide confidence
in the 
applicability of lattice QCD to predict other quantities of interest that might not be so well accessible experimentally.

The operators necessary to calculate nucleon structure require a renormalization
and matching to the continuum scheme used to analyse the experimental measurements.
The renormalization is performed non-perturbatively in the RI$^\prime$-MOM scheme
and then matched perturbatively to the continuum $\overline{MS}$ scheme to
compare to the phenomenological results for the moments of parton distribution functions.
In particular, we discuss the non-perturbative determination
of the renormalization factors
of the vector and axial vector operators with zero and one derivative.

Parton distribution functions (PDFs) are defined in Minkowski space and
therefore are not directly accessible to lattice calculations. However,
the moments of PDFs can be related to local operators that can be
calculated in Euclidean space with lattice methods.
The first moment of the parton distribution function, $q(x,\mu^2)$, is
given by
\begin{align}
\langle x \rangle_{q,\mu^2} = \int_{-1}^1\!\!dx\ x\: q(x,\mu^2) = \int_0^1\!\!dx\  x \left\{ q(x,\mu^2) + \overline{q}(x,\mu^2) \right\}.
\end{align}
We consider the isovector  combination $\left\langle x
\right\rangle_{u-d}$ in order to eliminate disconnected contributions.
It can be determined by evaluating the expectation value of the local
operator $O^{\mu\nu}$ given by
\begin{align}
\label{eqn:threepoint}
\left.\langle p, s|\underbrace{\overline{q}\gamma^{\left\{\mu\right.}
  iD^{\left.\nu\right\}} \tau^3 q}_{O^{\mu\nu}}
|p,s\rangle\right.\big|_{\mu^2} &= 2 \left\langle x
\right\rangle_{u-d,\mu^2} p^{\left\{\mu\right.}
p^{\left.\nu\right\}}. 
\end{align}
Here $q$ denotes the quark doublet $(u,d)$ and $\tau^3$ is the Pauli
matrix acting on the flavor indices. 
\section{Lattice techniques}
We use the Wilson twisted mass fermion action, which has the advantage of
leading to physical observables that are automatically $\mathcal{O}(a)$ improved
\cite{Frezzotti:2003ni}. Our $N_f=2+1+1$ setup, where we include
dynamical up, down, strange and charm quarks, maintains this automatic  $\mathcal{O}(a)$  improvement.
Thus no operator improvement is
necessary for $O^{\mu\nu}$ and $\left\langle x \right\rangle_{u-d}$ is accurate to $\mathcal{O}(a^2)$, thus providing an advantage
compared to the improved Wilson actions that would require additional calculations
for the operator improvement.
In Table~\ref{tab:details} we show the details of the
ensembles used so far in this work.  The pion masses range from $320$ to $450~\text{MeV}$. 
The volumes satisfy $m_\pi L > 4$ in order to keep finite
size effects  small.
The lattice
spacing for the coupling $\beta = 1.95$ has been estimated in the mesonic sector
to be $a\sim 0.078~\text{fm}$.  Further details are available in Ref.~\cite{Baron:2010bv}.

\begin{table}[ht]
\begin{center}
 \begin{tabular}[ht]{|l|l|l|l|l|l|l|l|}
  \hline
 $\beta$ & $L/a$ & $a\mu$ & $am_\pi$ &  $\left\langle x \right\rangle_{u-d}$ & stat. \\
  \hline
  1.95 & $32$ & 0.0075 & 0.18020(27)(3)  & 0.252(7)  & 412 \\
  1.95 & $32$ & 0.0055 & 0.15518(21)(33) & 0.240(8)  & 843 \\
  1.95 & $32$ & 0.0035 & 0.12602(30)(30) & 0.236(20) & 243 \\
  \hline
 \end{tabular}
\caption{\label{tab:details} In the first column we give
the bare coupling.  The second column is the spatial extent of the lattice and the temporal length is
always $T=2L$.  The bare quark mass in lattice units is given in the third column.  The fourth column is
the pion mass in lattice units with the 
statistical and systematic errors.  The results of $\langle x\rangle_{u-d}$ from
this work are given in the fifth column and the sixth lists the number of configurations
used in the calculation.} 
\end{center}
\vspace*{-0.5cm}
\end{table}
For the lattice calculation of $\langle x \rangle$ we need the 
evaluation of the proton three-point correlation function given by
\begin{align}
\label{eqn:threepoint_lat}
C_{3p}\left(t_\text{sink} - t_\text{source}, t_\text{operator} \right)
&= \text{Tr}\: \frac{1+\gamma_4}{2}
\sum\limits_{\mathbf{x_\text{operator}},\mathbf{x}_\text{sink}}
\left\langle  J_p(t_\text{sink},\mathbf{x}_\text{sink}
)O^{44}(x_\text{operator}) \bar{J}_p(x_\text{source})
\right\rangle,
\end{align}
where $\bar{J}_p$ ($J_p$) is an appropriate proton lattice creation
(annihilation) operator. 
The corresponding 2-point function is given by an identical expression
but with the omission of the operator insertion.
A standard ratio of the three-point function to the two-point 
function gives the desired matrix element 
in the large Euclidean-time limit:
 $t_\text{operator} - t_\text{source} \rightarrow \infty$ and
$t_\text{sink}     - t_\text{operator} \rightarrow \infty.$
\begin{figure}[htb]
\begin{minipage}{0.49\linewidth}
  \includegraphics[width=1.015\linewidth]{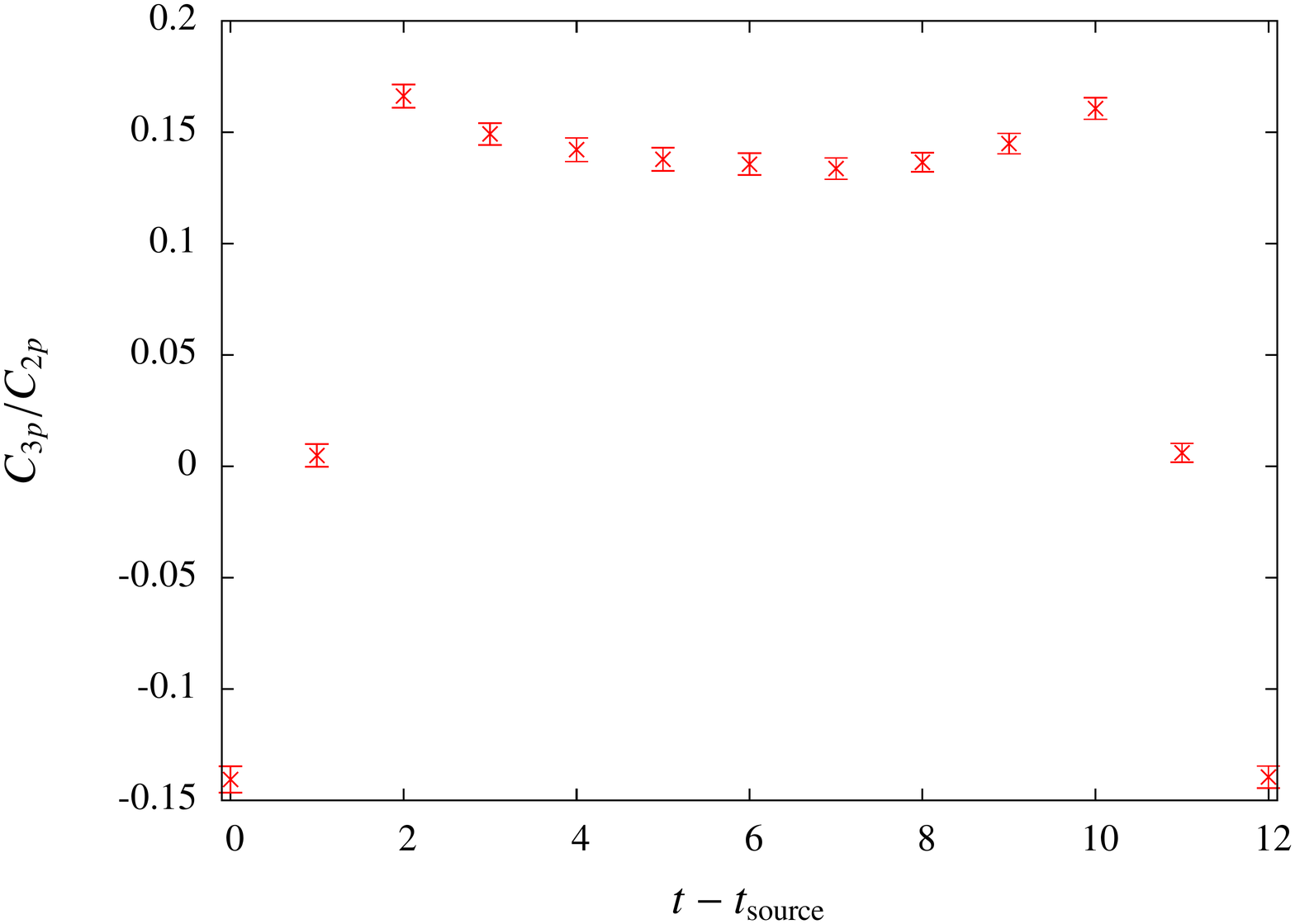}
  \caption{\label{fig:plateau} $C_{3p}/C_{2p}$ as a function of $t_\text{operator} - t_\text{source}$
          for fixed source-sink separation $t_\text{sink} - t_\text{source} = 12 a$ 
          on the ensemble with $m_\pi\approx 450~\mathrm{MeV}$.}
\end{minipage}\hspace{2mm}
\begin{minipage}{0.49\linewidth}
The ratio of $C_{3p}$ and $C_{2p}$
is computed as a function of $t_{\rm operator}$ and for large enough
time separations it becomes a constant yielding the matrix element of interest.
 This is illustrated
in Fig.~\ref{fig:plateau}, where we plot the dependence of the ratio
on $t_\text{operator}-
t_\text{source}$ for fixed source-sink separation $t_\text{sink} -
t_\text{source} = 12 a$. To evaluate $C_{3p}$ we construct sequential propagators at
the sink. In this approach one fixes the sink and source location, 
$x$ and $x_4^\prime$ respectively, as well as the quantum numbers of the
initial and final states. The sum 
over the spatial $\mathbf{x}^\prime$ can then be done implicitly by
performing an inversion with a source constructed
using the forward propagators generated at the source and the
state at the sink.
\end{minipage}
\end{figure}

The operator couples to the valence quarks
at an intermediate time $t_{\rm operator}$. 
Fixing the source-sink separation gives rise to contributions
from excited states and, therefore, the source-sink separation has to be chosen
sufficiently large to rule out those contributions but small enough
to avoid the exponentially
dropping signal-to-noise ratio.
It is essential to
use Gaussian smearing for the quarks in the proton interpolating 
 fields $J_p$  to increase the overlap of the creation and
annihilation operator with the ground state of the proton as well  APE smearing for
the gauge fields that enter the smearing functions to reduce noise.
 For the evaluation of the correlation functions we used the parallel
contraction code ahmidas \cite{Ahmidas}. 

\vspace*{-0.3cm}

\section{Results}

\vspace*{-0.3cm}

 \begin{figure}[ht]
\begin{minipage}[ht]{0.5\linewidth}
In order to assess cut-off effects
we examine the dependence on the lattice spacing, $a$, of the nucleon
mass which has been computed at three different values
of $a$. 
We show in Fig.~\ref{fig:m_N} the nucleon mass as a function
of $(a/r_0)^2$, where $r_0$ is the Sommer parameter.
 More details can be found in
Ref.~\cite{Drach:PoSLat2010}.
The observation is that  the
$\mathcal{O}(a^2)$ scaling is mild and compatible with
zero. Additionally the dependence of $\left\langle x \right\rangle_{u-d}$
on the lattice spacing that we have found using the $N_f=2$ ensembles
\cite{Alexandrou:2009qu, Alexandrou:plenary2010} is negligible.
Therefore, we have reason to believe that the lattice cut-off effects for
$\left\langle x \right\rangle_{u-d}$ using the $N_f=2+1+1$ will also be small. We
are currently investigating this issue.
\end{minipage}\hfill
\begin{minipage}{0.4\linewidth}
  \includegraphics[width=\linewidth]{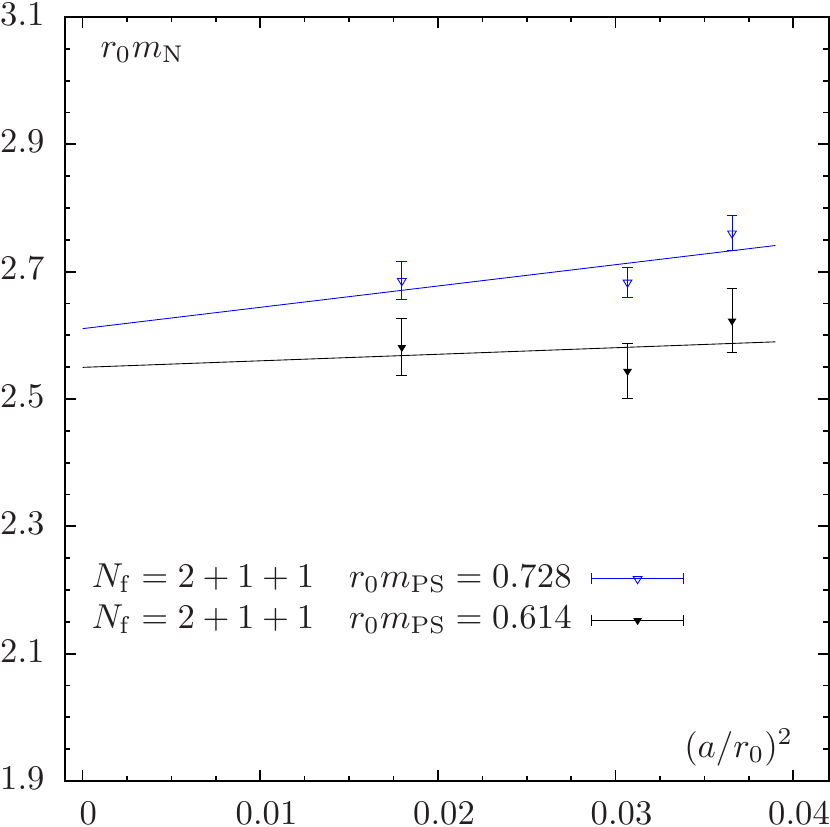} 
  \caption{\label{fig:m_N} The nucleon mass in units of $r_0$ as a
    function of $(a/r_0)^2$ for two reference pion masses.
}
\end{minipage}
 \end{figure}

The renormalization constants are computed non-perturbatively in the
RI$^\prime$-MOM scheme at different renormalization scales using the momentum source
method~\cite{Gockeler:1998ye}. The advantage of this method is a high
statistical accuracy and the evaluation of the vertex for any operator
including extended operators at no significant additional
computational cost. For the details of the non-perturbative
renormalization see Ref.~\cite{Alexandrou:2010}. 

In  the RI scheme the renormalization constants
are defined in the chiral limit. Since the mass of the strange
and charm quarks are fixed to their physical values in these
simulations, extrapolation to the chiral limit is not possible.
Therefore, in order to compute the renormalization constants needed
to obtain physical observations, ETMC has generated $N_f=4$ ensembles
at similar lattice spacings so that the chiral limit can be taken~\cite{Palao:PoSLat2010}.
We present  the renormalization factors for the local vector and axial vector
operators $Z^\mu_{\rm V}$ and $Z^\mu_{\rm A}$ as well as
  for the one-derivative vector and axial vector operators, 
$Z^{\mu\nu}_{\rm DV}$ and $Z^{\mu\nu}_{\rm DA}$, respectively. 
The later
fall into different irreducible representations of the hypercubic
group, depending on the choice of the external indices,
$\mu,\,\nu$. Hence, we distinguish between $Z_{\rm DV1}\,(Z_{\rm DA1})
= Z^{\mu\mu}_{\rm DV}\,\,(Z^{\mu\mu}_{\rm DA})$ and $Z_{\rm
  DV2}\,(Z_{\rm DA2}) = Z^{\mu\neq\nu}_{\rm DV}\,(Z^{\mu\neq\nu}_{\rm DA})$. 
\begin{figure}[ht]
\vspace{-3mm}
 \includegraphics[width=\linewidth, height=0.8\linewidth]{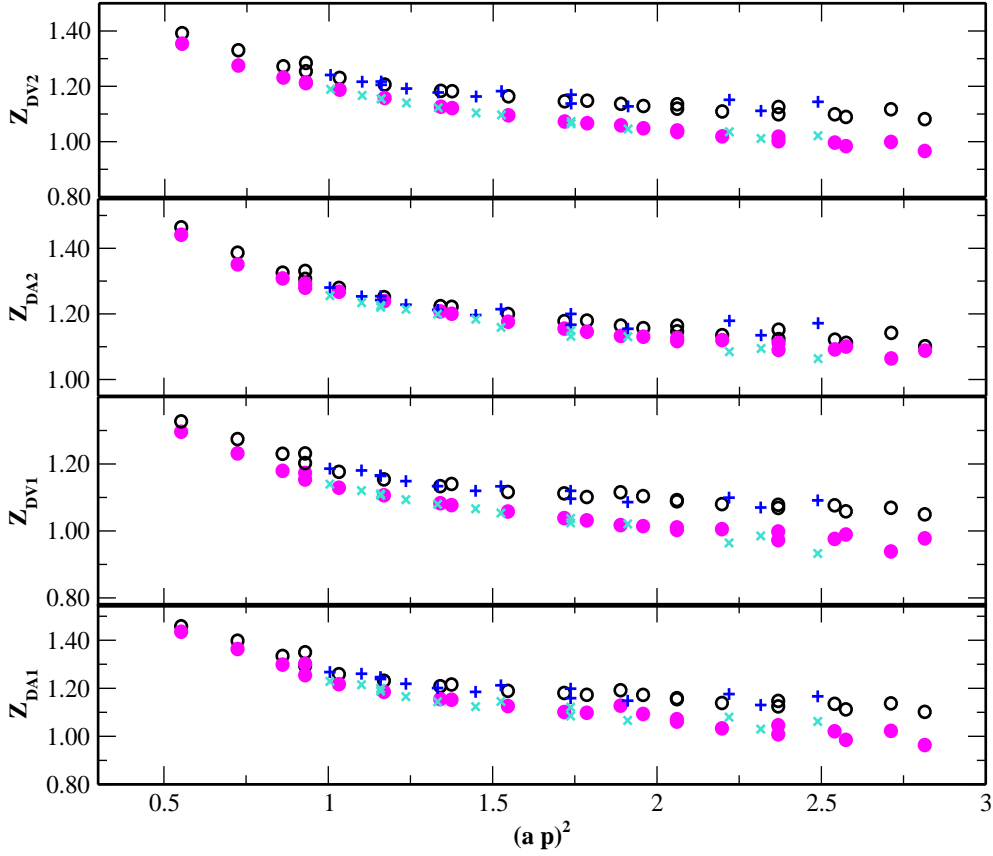}
  \caption{\label{fig:Z_RI} One-derivative renormalization constants
    as a function of the momentum squared in lattice units in the RI$'$-MOM scheme. We denote $N_f{=}4$ results by the open circles and $N_f{=}2{+}1{+}1$ by the crosses. Filled circles show the results after subtracting ${\cal  O}(a^2)$ terms for the $N_f=4$ case and x-symbols for the $N_f=2+1+1$ case.
}\vspace{-1mm}
 \end{figure}
Although we will use the
$N_f=4$ ensembles for the final determination of the
renormalization constants it is interesting
to  compute the renormalization constants also in the
$N_f=2+1+1$ theory and study their quark mass dependence.

\begin{figure}[ht]
\begin{minipage}{0.37\linewidth}
  \includegraphics[width=1.1\linewidth,height=0.7\linewidth]{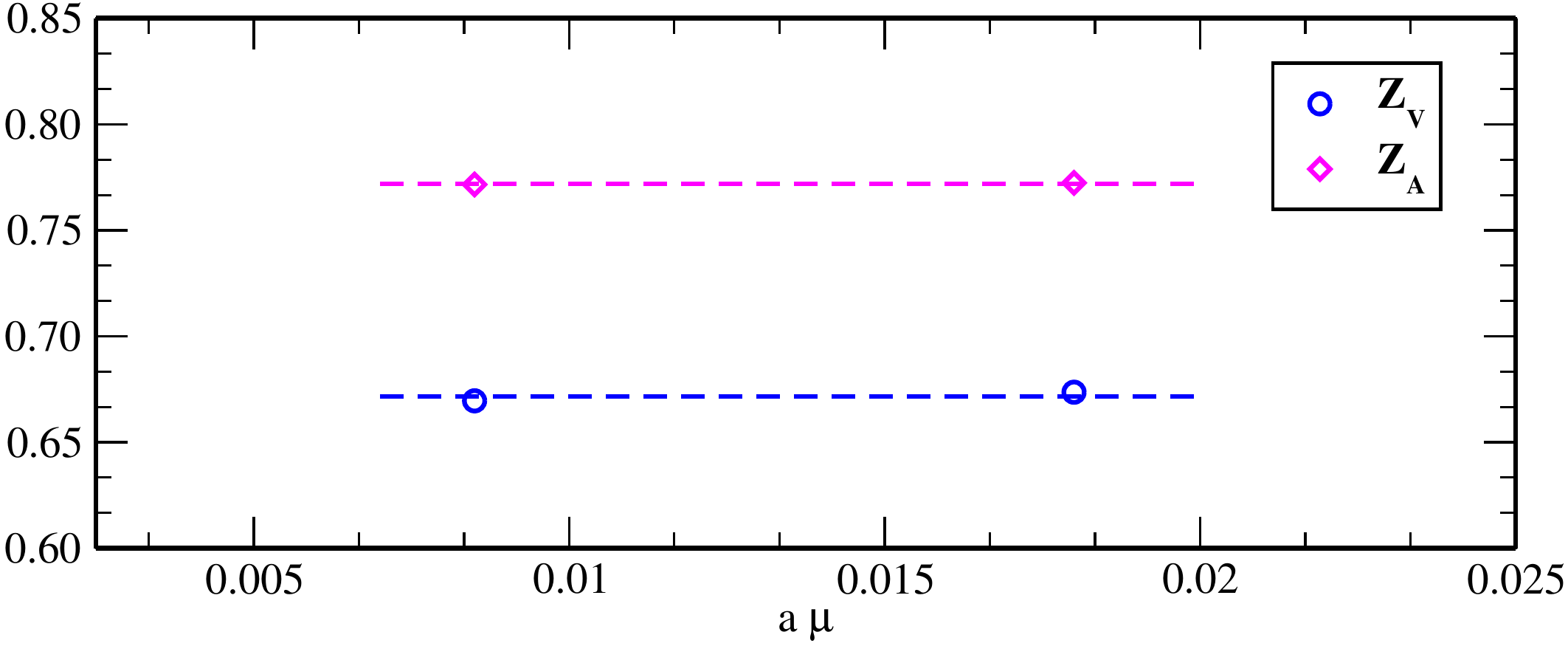} \\
\includegraphics[width=1.1\linewidth,height=0.7\linewidth]{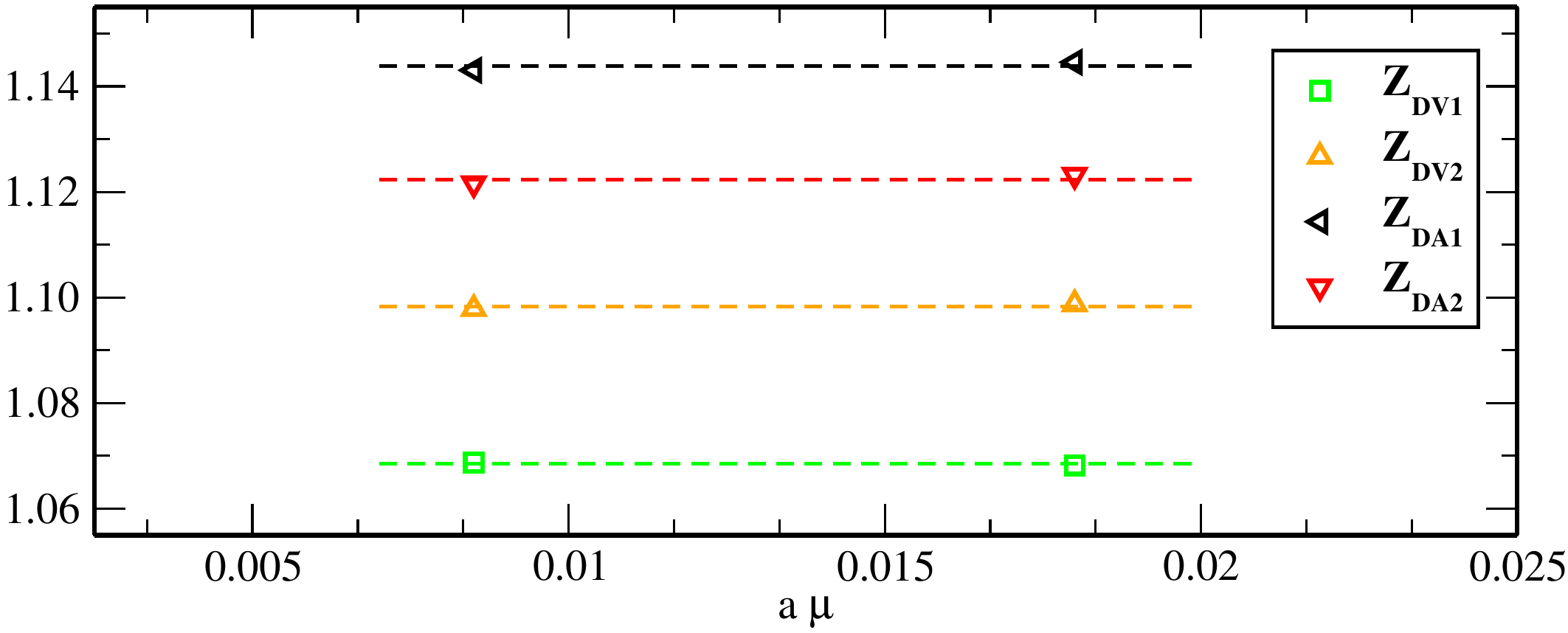} 
\vspace*{-0.5cm}
  \caption{\label{fig:Z_vs_mpi}$Z_V$ and $Z_A$ (upper panel) and renormalization constants 
   for the one-derivative operators (lower panel)
    as a function of the twisted mass. We have added 0.02 to the
value of $Z_{\rm DA1}$ to distinguish it from $Z_{\rm DA2}$.} 
\end{minipage}\hfill
\begin{minipage}{0.6\linewidth}
\hspace*{0.5cm}\includegraphics[width=1.1\linewidth]{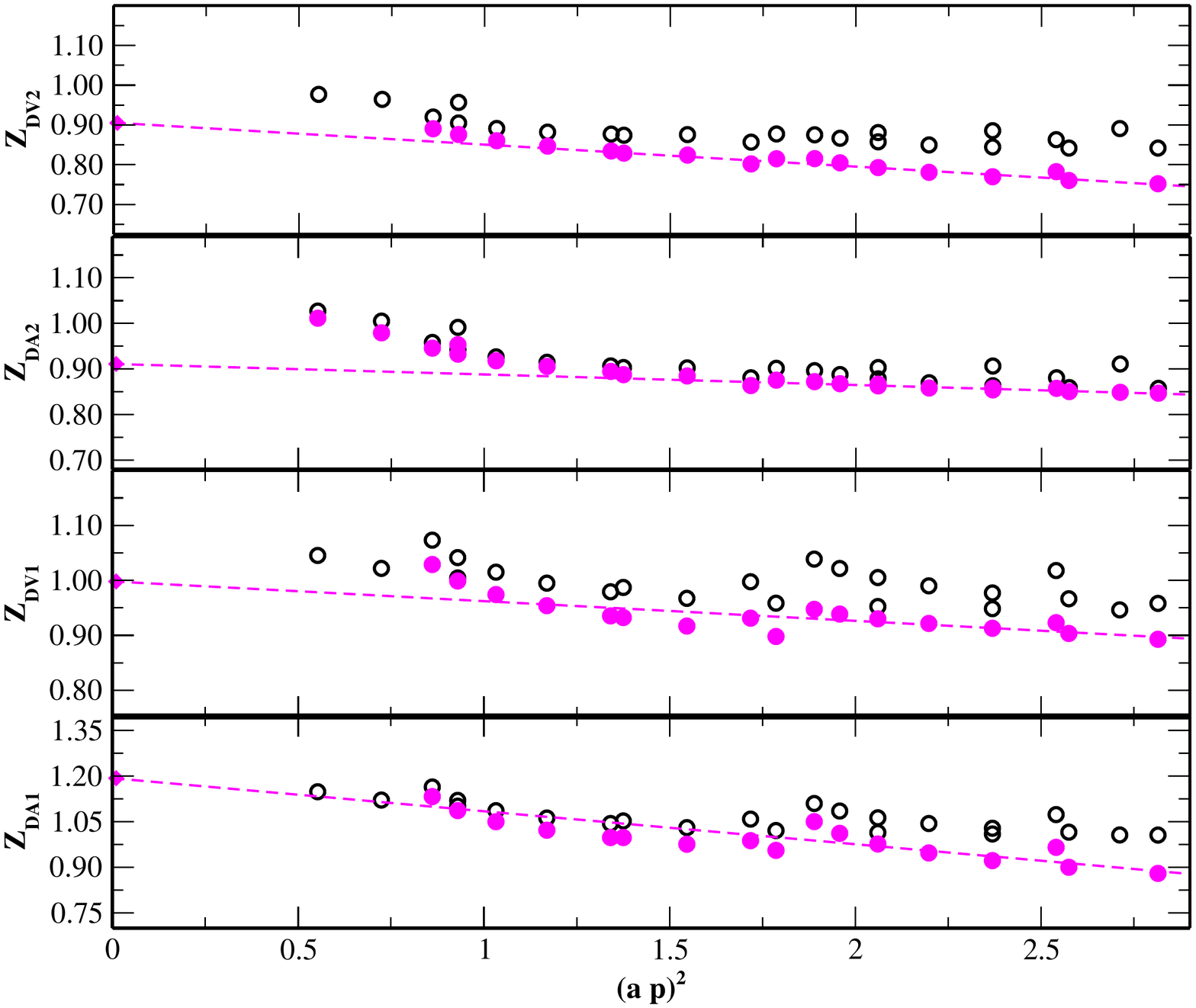}
\vspace*{-0.5cm}
  \hspace*{0.5cm}\caption{\label{fig:Z_MS}One-derivative renormalization constants
    as a function of the momentum squared in lattice units converted to the
      $\overline{MS}$-scheme for the $N_f=4$ ensemble.
Open circles denote results before perturbative subtraction of ${\cal  O}(a^2)$-terms 
and filled circles denote the ${\cal O}(a^2)$-subtracted results.}
\end{minipage}
\end{figure}

In Fig.~\ref{fig:Z_RI} we show results for 
both the $N_f=4$ and $N_f=2+1+1$ ensembles in the RI$^\prime$-MOM scheme.
As can be seen, we obtain compatible values for all the operators. The
same behavior is also observed in the case of $Z_V$ and $Z_A$.
This can be understood by examining the quark mass dependence of these results.
  In Fig.~\ref{fig:Z_vs_mpi} we show, for the $N_f=4$ case, the dependence of $Z_{\rm V},\,Z_{\rm
  A},\,Z_{\rm DV},\,Z_{\rm DA}$ on two light quark masses. The values
we find are consistent with each other. This explains the fact that 
the results in the
$N_f=4$ and $N=2+1+1$ cases are compatible. Furthermore it makes any
 extrapolation of $N_F=4$ results to the chiral limit straight forward.

The renormalization scale dependent $Z_{\rm DV}$ and $Z_{\rm DA}$ need
to be converted to the continuum ${\overline{\rm MS}}$-scheme, and
for this we use a conversion factor computed in perturbation theory to
${\cal O}(g^6)$. They are also evolved perturbatively to a reference
scale, which is chosen to be (2~GeV)$^2$. The constant that
renormalizes $\left\langle x \right\rangle_{u-d}$ is $Z_{\rm
  DV1}$. 
The results are shown in Fig.~\ref{fig:Z_MS} both before subtracting
the perturbative ${\cal  O}(a^2)$-terms and after. Using the subtracted
results, the preliminary value of $Z_{\rm DV1}$ in the $\overline{\rm
  MS}$-scheme at (2~GeV)$^2$ is $Z_{\rm DV1}=0.998$ calculated using the
$N_f=4$ ensemble with $\beta=1.95$ and $am_\pi=0.194$.

 \begin{figure}[ht]
\begin{center}
  \includegraphics[width=0.7\textwidth]{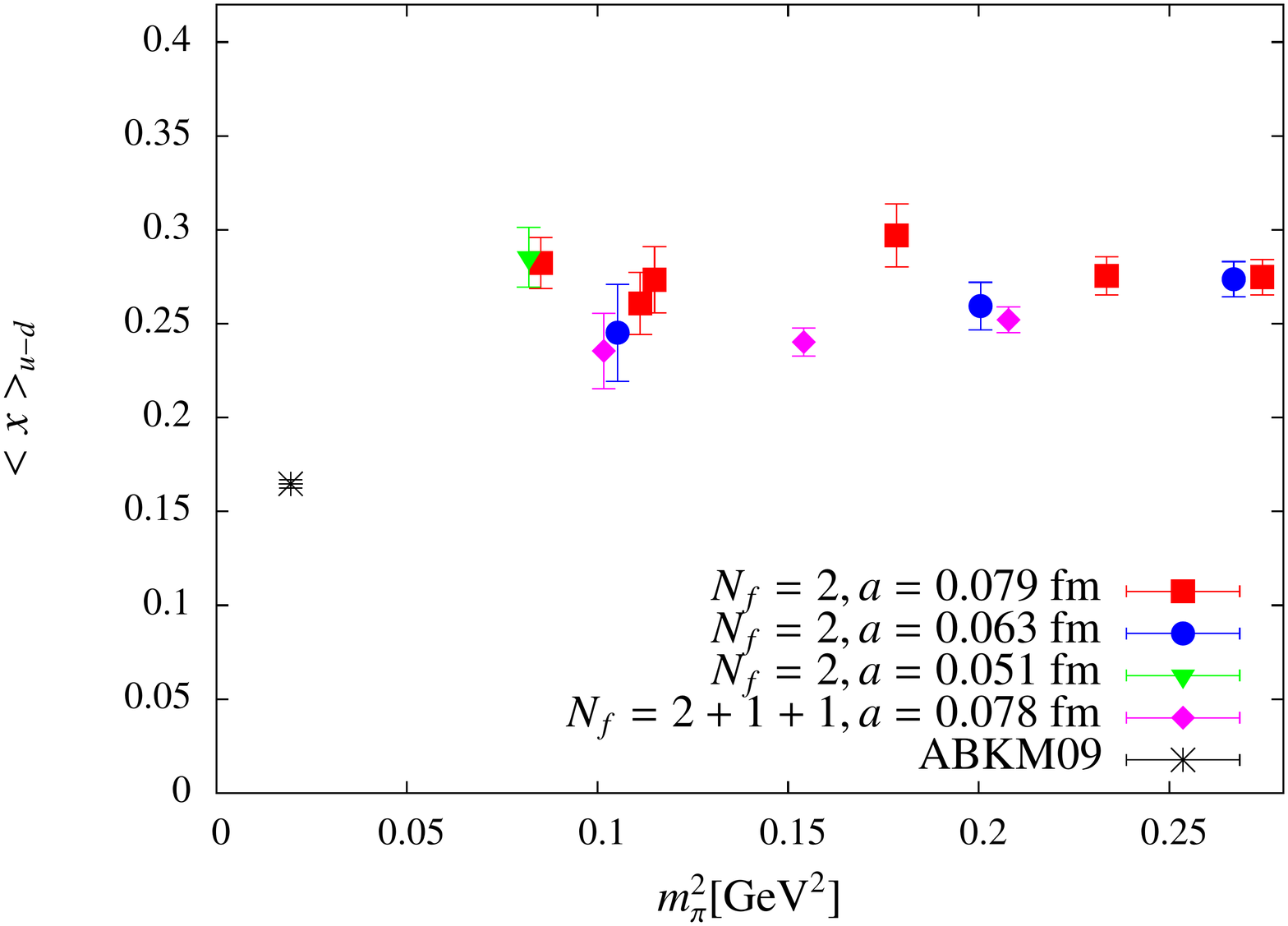}
  \caption{\label{fig:avX_Nf2and2p1p1} $\left\langle x \right\rangle_{u-d}$ obtained from
           $N_f=2$ and from $N_f=2+1+1$ ensembles for a range of pion masses.
           The experimental value was taken from the
           ABKM09 analysis~\cite{Alekhin:2009ni}.}
\end{center}
\end{figure}

In Fig.~\ref{fig:avX_Nf2and2p1p1} we show the results
for $\left\langle x \right\rangle_{u-d}$ for several pion masses
 for the $N_f=2+1+1$ ensembles and give the values in Table~\ref{tab:details}.
We compare with the $N_f=2$ results for similar pion 
masses~\cite{Alexandrou:201X}.
As can be seen, the $N_f=2$ and $N_f=2+1+1$ results are consistent showing
that the effects of the strange and charm quarks in the sea are
small compared to the, admittedly large, statistical errors. The results do not show a strong mass 
dependence and are higher than experiment.
This behavior is consistent with the
results of other collaborations~\cite{Alexandrou:plenary2010,Renner:2010ks}.
Thus calculations at even lower pion masses are needed to
shed light on the behaviour closer to the physical point.
\section{Conclusions}
We presented results on $\left\langle x \right\rangle_{u-d}$
with $N_f=2+1+1$  twisted mass fermions for pion masses in the
range of 320~MeV to 450~MeV at  $\beta=1.95$ ($a\sim0.078\,\text{fm}$).
We used non-perturbative renormalization calculated for an $N_f=4$ ensemble
at the same coupling. The results are in agreement
with those obtained with  $N_f=2$ twisted mass fermions.
As expected from $N_f=2$, there is no strong mass dependence and
the results are higher than experiment for the pion masses used in these calculations.
The chiral behaviour of $\left\langle x
\right\rangle_{u-d}$ will be studied using smaller pion masses and
paying particular attention to excited state contributions in the
three-point function.
\section*{Acknowledgments}
This work is supported in part by  the DFG
Sonder\-for\-schungs\-be\-reich/ Trans\-regio SFB/TR9 and by funding received from the
 Cyprus Research Promotion Foundation under contracts EPYAN/0506/08,
KY-$\Gamma$/0907/11/ and TECHNOLOGY/$\Theta$E$\Pi$I$\Sigma$/0308(BE)/17. 
It is additionally coauthored in part by Jefferson 
Science Associates, LLC under U.S. DOE Contract No. DE-AC05-06OR23177.
This work was performed using HPC ressource from GENCI/IDRIS (Grant 2010/052271).


\begin{thebibliography}{99}

\bibitem{Baron:2010bv}
  R.~Baron {\it et al.},
  JHEP {\bf 1006} (2010) 111
  [arXiv:1004.5284 [hep-lat]].

\bibitem{Baron:2009zq}
  R.~Baron {\it et al.},
  PoS {\bf Lattice 2009} (2009) 104
  [arXiv:0911.5244 [hep-lat]].

\bibitem{Frezzotti:2003ni}
  R.~Frezzotti and G.~C.~Rossi,
  JHEP {\bf 0408} (2004) 007
  [arXiv:hep-lat/0306014].

\bibitem{Ahmidas}
  A.~Deuzeman, S.~Dinter, S.~Reker,
  Ahmidas -- Parallelized contraction codes for lattice QCD
  \textit{http://code.google.com/p/ahmidas}


\bibitem{Drach:PoSLat2010}
  V.~Drach {\it et al.}  [ETM Collaboration],
  PoS {\bf Lattice 2010}, 101 (2010).


\bibitem{Alexandrou:2009qu}
  C.~Alexandrou {\it et al.}  [ETM Collaboration],
  Phys.\ Rev.\  D {\bf 80} (2009) 114503
  [arXiv:0910.2419 [hep-lat]].

\bibitem{Alexandrou:plenary2010} C. Alexandrou, Plenary talk at Lattice 2010,
PoS {\bf Lattice 2010}, 001 (2010).

\bibitem{Gockeler:1998ye}
M. G\"ockeler, R. Horsley, H. Oelrich, H. Perlt, D. Petters, P.E.L. Rakow, A. Schafer, G. Schierholz, A. Schiller,
Nucl. Phys. {\bf B544} (1999) 699, [{\tt hep-lat/9807044}].

\bibitem{Alexandrou:2010}
C. Alexandrou, M. Constantinou, T. Korzec, H. Panagopoulos,
F. Stylianou, arXiv:1006.1920.


\bibitem{Palao:PoSLat2010}
  D.~Palao {\it et al.}  [ETM Collaboration],
Lattice 2010, in preparation.

\bibitem{Alexandrou:201X}
  C.~Alexandrou {\it et al.}  [ETM Collaboration], in preparation.

\bibitem{Renner:2010ks}
  D.~B.~Renner,
  PoS {\bf LAT2009}, 018 (2009), [arXiv:1002.0925 [hep-lat]].

\bibitem{Alekhin:2009ni}
 S. Alekhin, J. Blumlein, S. Klein, S. Moch,
 Phys. Rev. D81(2010) 014032, [arXiv:0908.2766 [hep-ph]].


\end{thebibliography}
\end{document}